\documentclass{PoS}

\title{Complementarity of Forward-Backward Asymmetry for discovery of $Z^\prime$ bosons at the Large Hadron Collider}

\ShortTitle{$A_{FB}$ for $Z^\prime$ bosons at the LHC}

\author{\speaker{Juri Fiaschi}\\
        School of Physics \& Astronomy, University of Southampton\\
        E-mail: \email{juri.fiaschi@soton.ac.uk}}

\author{Elena Accomando\\
       School of Physics \& Astronomy, University of Southampton\\
       E-mail: \email{e.accomando@soton.ac.uk}}
       
\author{Alexander Belyaev\\
       School of Physics \& Astronomy, University of Southampton\\
       E-mail: \email{a.belyaev@soton.ac.uk}}
       
\author{Ken Mimasu\\
       School of Physics \& Astronomy, University of Sussex\\
       E-mail: \email{k.mimasu@sussex.ac.uk}}
       
\author{Stefano Moretti\\
       School of Physics \& Astronomy, University of Southampton\\
       E-mail: \email{s.moretti@soton.ac.uk}}

\author{Claire H. Shepherd-Themistocleous\\
       Particle Physics Department, Rutherford Appleton Laboratory\\
       E-mail: \email{claire.shepherd@stfc.ac.uk}}
       
\abstract{The Forward-Backward Asymmetry (AFB) in $Z^\prime$ physics is commonly only thought of as an observable which possibly allows one to 
profiling  a $Z^\prime$ signal by distinguishing different models embedding such (heavy) spin-1 bosons.
In this brief review, we examine the potential of AFB in setting bounds on or even discovering a $Z^\prime$ at the Large Hadron Collider (LHC) and proof  
that it might be a powerful tool for this purpose.
We analyse two different scenarios: $Z^\prime$s with a narrow and wide width, respectively. We find that, in both cases, AFB can complement the 
conventional searches in accessing $Z^\prime$ signals traditionally based on cross section measurements only.}

\FullConference{The European Physical Society Conference on High Energy Physics\\
		22--29 July 2015\\
		Vienna, Austria}

\begin{document}
 
 \section{Introduction}
Extra gauge bosons are present in many Beyond Standard Model (BSM) theories. Phenomenologically the simplest possible scenario appears adding an extra $U(1)$ symmetry to the SM gauge group. Following this approach we have been able to study the three main classes of models that predict a $Z^\prime$: $E_6$, Generalized Left-Right ($GLR$) symmetric and Generalized Standard Model ($GSM$) \cite{Accomando:2010fz}.
All these classes predict rather narrow $Z^\prime$s ($\Gamma_{Z^\prime} / M_{Z^\prime} \sim 0.5 - 12 \%$).

Experimental searches optimized for such narrow resonances assume a very visible peak with a Breit-Wigner line-shape over the SM background, when looking at the cross section sampled over the invariant mass of the $Z^\prime$ decay products.
On this basis, the 95\% Confidence Level (C.L.) upper bound on the cross section is derived and limits on the mass of the resonance are extracted within the above benchmark models.
Theoretically, predictions are mostly calculated in the Narrow Width Approximation (NWA), although occasionally including Finite Width (FW) and interference effects. Putting an appropriate cut on in the invariant mass spectrum, these contribution can be kept under control (below 10\%) in a model independent way~\cite{Accomando:2013sfa}. 

However, there exist many scenarios where the NWA is not properly valid. Composite Higgs Models scenarios where the $Z^\prime$ is differently coupled to the first two fermion generations with respect to the third one or where the new gauge sector mixes with the SM neutral one, are frameworks where broad  $Z^\prime$s are possible.
Here, the ratio $\Gamma_{Z^\prime}/M_{Z^\prime}$ can reach 50\% or more.

Experimental searches studying these ``effectively'' non-resonant cases are essentially counting experiments: an integration over the overall invariant mass spectrum beyond a minimum invariant mass  seeks an excess of events spread over the SM background.
The analysis, even if improved using optimized kinematical cuts, still maintains some fragile aspects as it relies on the good understanding of the SM background, especially in the control region which could be contaminated by BSM physics and could affect the $Z^\prime$ decay products invariant mass distribution in the low mass region.
This effect is increased in the case of wide resonances, that is why their search strategies turn out to be quite problematic.

In this article we study the effects of the inclusion of another observable into the analysis of heavy neutral resonances: the Forward-Backward Asymmetry (AFB). We explore the complementary potential of AFB with respect to the ``bump'' 
or ``counting experiment'' searches in both the narrow and wide  $Z^\prime$ framework, respectively. Note that,
in current literature, this observable is usually adopted as a post-discovery tool to interpret the  evidence of a peaked signal and to possibly disentangle between different theoretical models that would predict it.
Our purpose is to show that AFB can also be used in the very same search process as a discovery observable (see Ref. \cite{Accomando:2015cfa}).
We focus on the golden channel for $Z^\prime$  searches at the LHC, {\it i.e.}, the Drell-Yan (DY) process $pp\rightarrow l^+l^-$ with $l=e, \mu$.

The article is organized as follows. 
In sect. 2 we derive current and projected bounds for $Z^\prime$ model benchmarks for the LHC at 8 and 13 TeV.
In sect. 3 we discuss the role of AFB  in the context of either narrow or wide resonance searches.
In sect. 4 we summarize and conclude.

\section{Bounds on the $Z^\prime$ mass}
We validated our analysis reproducing the current experimental limits obtained by, \textit{e.g.}, the CMS collaboration after the 7 and 8 TeV runs with about 20 $fb^{-1}$ of luminosity \cite{Khachatryan:2014fba}.
These limits are computed through the ratio $R_\sigma = \sigma (pp\rightarrow Z^\prime\rightarrow l^+l^-)/\sigma (pp\rightarrow Z, \gamma \rightarrow l^+l^-)$.
\ Here $R_\sigma$ has been calculated at the Next-to-Next-to-Leading Order (NNLO) in QCD. 
The resulting exclusion limits we compute include FW and interference effects. The values we obtain are summarized in Tab. 1: they match the reported limits by CMS for the benchmark models $GSM-SSM$ and $E_6-\chi$ within the accuracy of 1-2 \%.
These models with an extra $U(1)$  group  predict intrinsically narrow $Z^\prime$s and in this context we expect FW and interference effects to be small. This is why we have been able to match CMS results with great precision even though their results are calculated in NWA.

\begin{table}
  \label{tab:events8tev}
  \begin{tabular}{|c||c|c|c|c|c|c|}
    \hline
      Class & \multicolumn{6}{c|}{$E_6$} \\
    \hline
      $U^\prime (1)$ Models  & $\chi$ & $\phi$ & $\eta$ & $S$ & $I$ & $N$ \\
    \hline
      $M_{Z^\prime}$ [GeV] & 2700 & 2560 & 2620 & 2640 & 2600 & 2570 \\
    \hline
  \end{tabular}
  \\
  \begin{tabular}{|c||c|c|c|c|c|c|c|}
    \hline
      Class & \multicolumn{4}{c|}{$GLR$} & \multicolumn{3}{c|}{$GSM$} \\
    \hline
      $U^\prime (1)$ Models  & $R$ & $B-L$ & $LR$ & $Y$ & $SSM$ & $T_{3L}$ & $Q$ \\
    \hline
      $M_{Z^\prime}$ [GeV] & 3040 & 2950 & 2765 & 3260 & 2900 & 3135 & 3720\\
    \hline
  \end{tabular}
\caption{Bounds on the $Z^\prime$ mass we have derived from the latest direct searches data performed by CMS at the 7 and 8 TeV LHC with integrated luminosity $\mathcal{L}=20 fb^{-1}$. They are in good agreement with the latest CMS published results~\cite{Khachatryan:2014fba}.}
\end{table}


After having verified the reliability of our code, we have been able to project future discovery and exclusion limits for the next run of the LHC at 13 TeV and with a luminosity of 300 $fb^{-1}$. In both cases we have taken into account the published acceptance $\times$ efficiency corrections and a Poisson statistic approach has been used for computing the significance of the signal.
Requiring for the latter a significance of 2 for exclusion and of 5 for discovery, we obtain the results summarized in Tab. \ref{tab:events13tev}. It is worth to stress that the significances we are computing here are purely statistical and they do not include systematic uncertainties.

\begin{table}
  \label{tab:events13tev}
  \begin{tabular}{|c||c|c|c|c|c|c|}
    \hline
      Class & \multicolumn{6}{c|}{$E_6$} \\
    \hline
      $U^\prime (1)$ Models  & $\chi$ & $\phi$ & $\eta$ & $S$ & $I$ & $N$ \\
    \hline
      $M_{Z^\prime}$ [GeV] & 4535 & 4270 & 4385 & 4405 & 4325 & 4290 \\
    \hline
      $M_{Z^\prime}$ [GeV] & 5330 & 5150 & 5275 & 5150 & 5055 & 5125 \\
    \hline
  \end{tabular}
  \\
  \begin{tabular}{|c||c|c|c|c|c|c|c|}
    \hline
      Class & \multicolumn{4}{c|}{GLR} & \multicolumn{3}{c|}{GSM} \\
    \hline
      $U^\prime (1)$ Models  & $R$ & $B-L$ & $LR$ & $Y$ & $SSM$ & $T_{3L}$ & $Q$ \\
    \hline
      $M_{Z^\prime}$ [GeV] & 5175 & 5005 & 4655 & 5585 & 4950 & 5340 & 6360 \\
      \hline
      $M_{Z^\prime}$ [GeV] & 6020 & 5855 & 5495 & 6435 & 5750 & 6180 & 8835 \\
    \hline
  \end{tabular}
\caption{Projection of discovery limits (first row) and exclusion limits (second row) on the $Z^\prime$ mass from direct searches at the forthcoming Run II of the LHC at 13 TeV. We assume $\mathcal{L}=300 ~fb^{-1}$.}
\end{table}

\section{The role of AFB in $Z^\prime$ searches: narrow and wide heavy resonances}

We define $A_{FB}^*$ as follows:
\vspace*{-0.25truecm}
\begin{equation}
\label{eq:AFB}
\frac{d\sigma}{d\cos\theta_l^*}\propto \sum_{spin,col}\left|\sum_i\mathcal{M}_i\right|^2=\frac{\hat{s}^2}{3}\sum_{i,j}|P^*_iP_j|[(1+\cos^2\theta_l^*)C^{ij}_S+2\cos\theta_l^* C^{ij}_A] 
\end{equation}

\noindent where $\theta_l^*$ is the lepton angle with respect to the quark direction in the di-lepton Centre-of-Mass (CM) frame, which can be derived from the measured four-momenta of the di-lepton system in the laboratory frame. 
The AFB is indeed given by the coefficient of the contribution to the angular distribution linear in $\cos\theta_l^*$. In Eq. (\ref{eq:AFB}), $\sqrt{\hat{s}}$ is the invariant mass of the di-lepton system and $P_i$ and $P_j$ are the propagators of the gauge bosons involved in the process. 
At the tree-level, DY production of charged lepton pairs is mediated by three gauge bosons: the SM photon and $Z$-boson and the hypothetical $Z^\prime$. These three vector boson exchanges all participate in the matrix element squared. The interferences amongst these three particles  have to be take into account properly.
Finally, the factors $C_{S}^{ij}$ and $C_{A}^{ij}$ in the angular distribution given in Eq. (\ref{eq:AFB}) are the parity symmetric and anti-symmetric coefficients function of the  quark and lepton chiral couplings, $q_{L/R}^i$ and $e_{L/R}^i$, to the $i$-boson with $i=\{\gamma, Z, Z^\prime\}$:
%
$C_{S}^{ij}=(q_L^i q_L^j+ q_R^i q_R^j)(e_L^i e_L^j+ e_R^i e_R^j),~
C_{A}^{ij}=(q_L^i q_L^j- q_R^i q_R^j)(e_L^i e_L^j- e_R^i e_R^j)$.

Looking at these expressions it is clear that the analysis of AFB can give us complementary information with respect to the cross section distribution (which is proportional to the sum of the squared chiral couplings)
about the couplings between the $Z^\prime$ and the fermions.
This feature has motivated several authors to study the potential of AFB  in interpreting a possible $Z^\prime$ discovery obtained in the usual cross section hunt.

The AFB is obtained by integrating the lepton angular distribution forward and backward with respect to the quark direction.This can also be done as a function of $\sqrt{\hat{s}}\equiv M_{ll}$, thereby introducing a differential AFB. 
As in $pp$ collisions the original quark direction is not known, one has to extract it from the kinematics of the di-lepton system. In this analysis, we follow the criteria of Ref. \cite{Dittmar:1996my} and simulate the quark direction from the boost of the di-lepton system with respect to the beam axis ($z$-axis).

In the following we are going to show the impact of AFB on the significance of the signal. For this purpose we give the general definition of significance $\alpha$ for a generic observable:
\vspace*{-0.20truecm}
\begin{equation}
\alpha=\frac{|O_1-O_2|}{\sqrt{\delta O^2_1+\delta O^2_2}},
\end{equation}

\noindent where the $O_i$s ($i=1,2$) are the value of the observable in two hypothesis scenarios with uncertainty $\delta O_i$. In the case of AFB we will use the statistical uncertainty:
\vspace*{-0.20truecm}
\begin{equation}
\delta A_{FB}^\ast =\sqrt{\frac{4}{\mathcal{L}}\frac{\sigma_{F}\sigma_{B}}{(\sigma_{F}+\sigma_{B})^3}} = 
\sqrt{\frac{(1-A_{FB}^{\ast2})}{\sigma\mathcal{L}}} = \sqrt{\frac{(1-A_{FB}^{\ast2})}{N}},
\end{equation}

\noindent where $\mathcal{L}$ is the integrated luminosity and $N$ the total number of events.
Since the significance is proportional to the root of the total number of events, the imposition of a stringent cut on the boost variable, $y_{l\bar{l}}$, in spite of guiding the AFB spectrum towards its true line shape, will decrease the statistics and, by consequence, the resulting significance of the signal. 

For this reason in the following we are not going to impose any rapidity cut and we are going to show how AFB can be used also as a powerful tool to search for new physics.

\subsection{Narrow heavy resonances}

\begin{figure}
\centering
\includegraphics[width=0.47\textwidth]{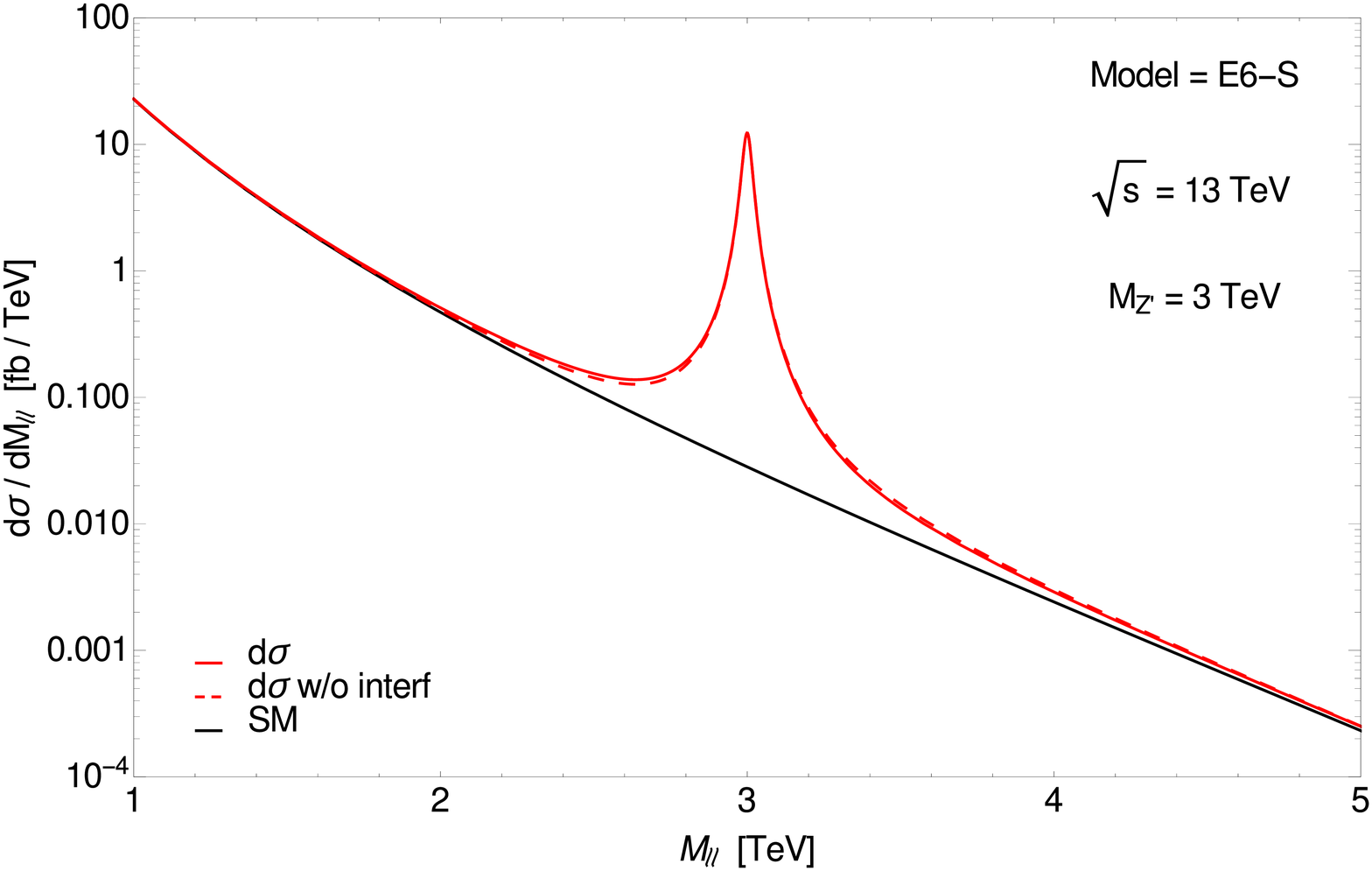}
\centering
\includegraphics[width=0.47\textwidth]{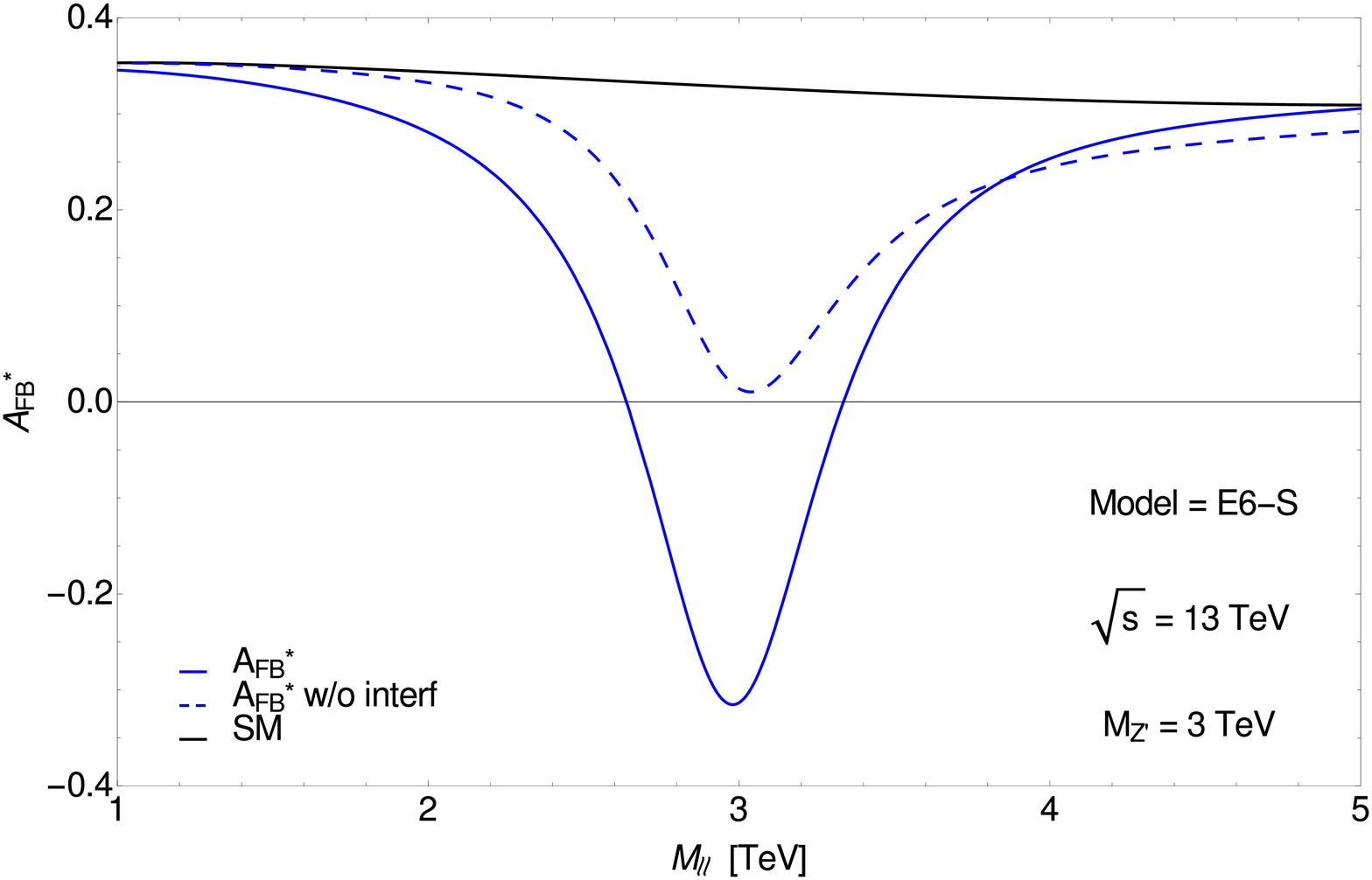}
\caption{Hypothetical signal in the cross section (Left) and $A_{FB}^*$ (Right) distributions produced by a $Z^\prime$ with mass $M_ {Z^\prime}$ = 3 TeV, as predicted by the $E_6-S$ model, at the LHC with  $\sqrt{s}$ = 13 TeV. No cut on the di-lepton rapidity is imposed: $|y_{l\bar{l}}|\ge 0$.}
\label{fig:S_model}
\end{figure}

We start comparing the shape of the AFB distribution as a function of the di-lepton invariant mass $\sqrt{\hat{s}}\equiv M_{ll}$.
We are showing here the case of the $E_6-S$ model (Fig. \ref{fig:S_model}).

As one can see, the role played by the interference is extremely important. In this case for instance the AFB peak is strongly accentuated by interference effects.
In contrast, the cross section distribution is almost interference free if the $|M_{l\bar l}-M_{Z^\prime}|\le 0.05\times E_{\rm{LHC}}$ cut is imposed
\cite{Accomando:2013sfa}. In interpreting the experimental data coming from AFB measurements  it is therefore mandatory to include the interference independently of any kinematical cut.

\begin{figure}
\centering
\includegraphics[width=0.47\textwidth]{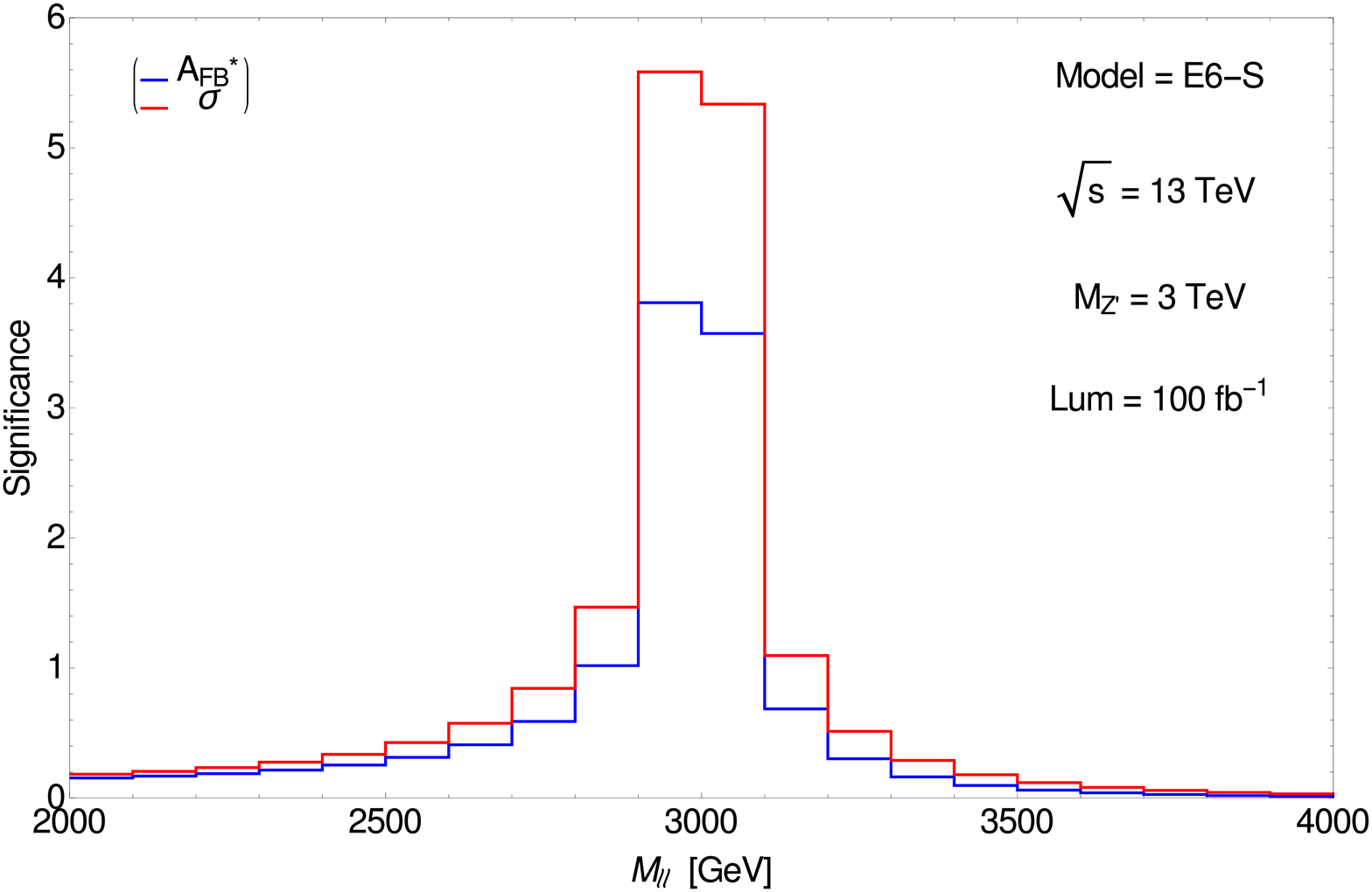}
\caption{Binned significance of an hypothetical signal produced by a $Z^\prime$ with mass $M_ {Z^\prime}$ = 3 TeV, as predicted by the $E_6-S$ model, at the LHC with  $\sqrt{s}$ = 13 TeV and $\mathcal{L}=100~fb^{-1}$, for the two observables: cross section and $A_{FB}^*$.}
\label{fig:S_significance}
\end{figure}

In terms of significance of the signal (Fig. \ref{fig:S_significance}), in the $E_6-S$ case we find that what we get from $\sigma$  is comparable with what we get from  $A_{FB}^*$.
This means that 
the latter can be used as a valid alternative to the fomer, also because it  is very reliable in terms of systematic uncertainties: since it comes from the ratio of cross sections, strong cancellations happen between the (similar) uncertainties in the forward and backward directions, upon taking into account their mutual correlations.

\subsection{Wide heavy resonances}
Next, we discuss the role of  $A_{FB}^*$ in searches for a new $Z^\prime$ characterized by a large width. 
Such a wide particle is predicted by various scenarios, like strongly interacting and/or non-universal models.
Herein we will consider an example of the latter, well described in the literature \cite{Malkawi:1999sa, Kim:2014afa}. The large parameter space of the model
allows one to explore the phenomenology of a wide $Z^\prime$ boson. Indeed we are going to consider the case in which the $Z^\prime$ is more coupled with the
fermions third generation, resulting in a quite large resonance width, and at the same time satisfying current exclusion limits due to the weaker coupling with 
electrons and muons.

\begin{figure}
\centering
\includegraphics[width=0.47\textwidth]{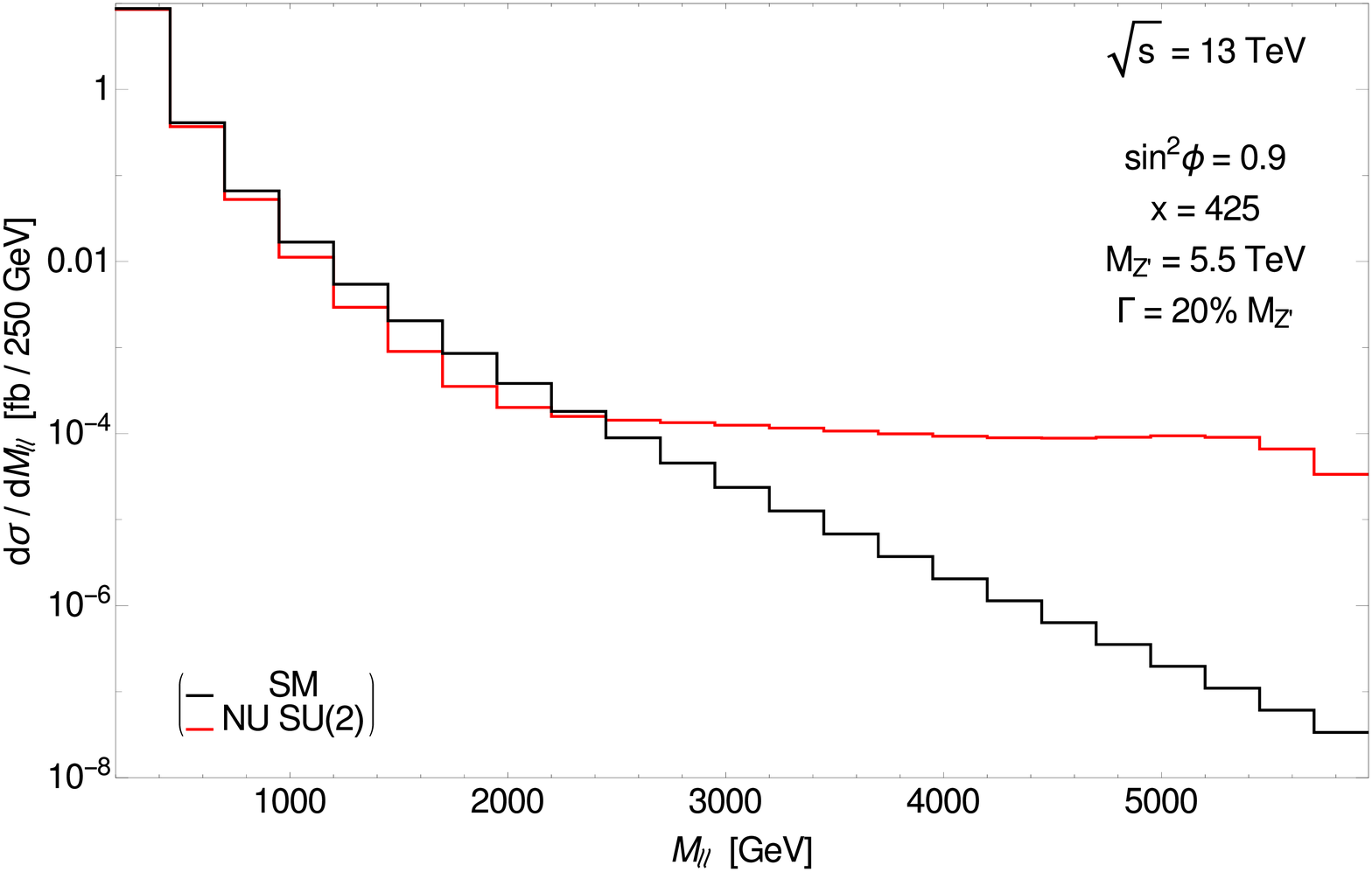}
\centering
\includegraphics[width=0.47\textwidth]{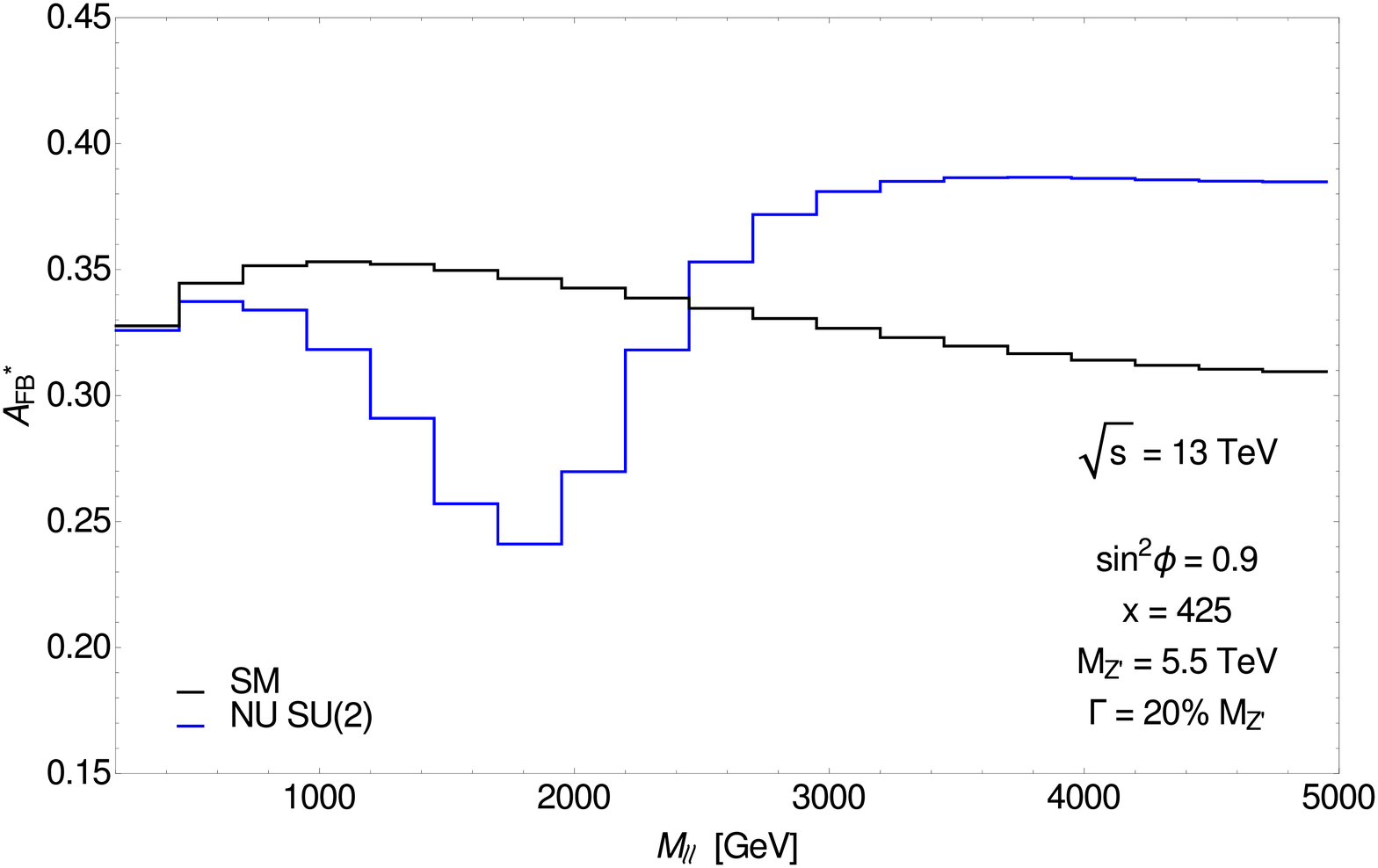}
\caption{Binned differential cross section (Left) and $A_{FB}^*$ (Right) distributions as a function of the di-lepton invariant mass as predicted by a Non-Universal $SU(2)$ model for a $Z^\prime$ with mass $M_ {Z^\prime}$ = 5.5 TeV and $\Gamma_{Z^\prime}/M_{Z^\prime}=20\%$. 
The results are for the LHC at $\sqrt{s}$ = 13 TeV and no rapidity cuts are applied.}
\label{fig:NUSU2}
\end{figure}

\begin{figure}
\centering
\includegraphics[width=0.47\textwidth]{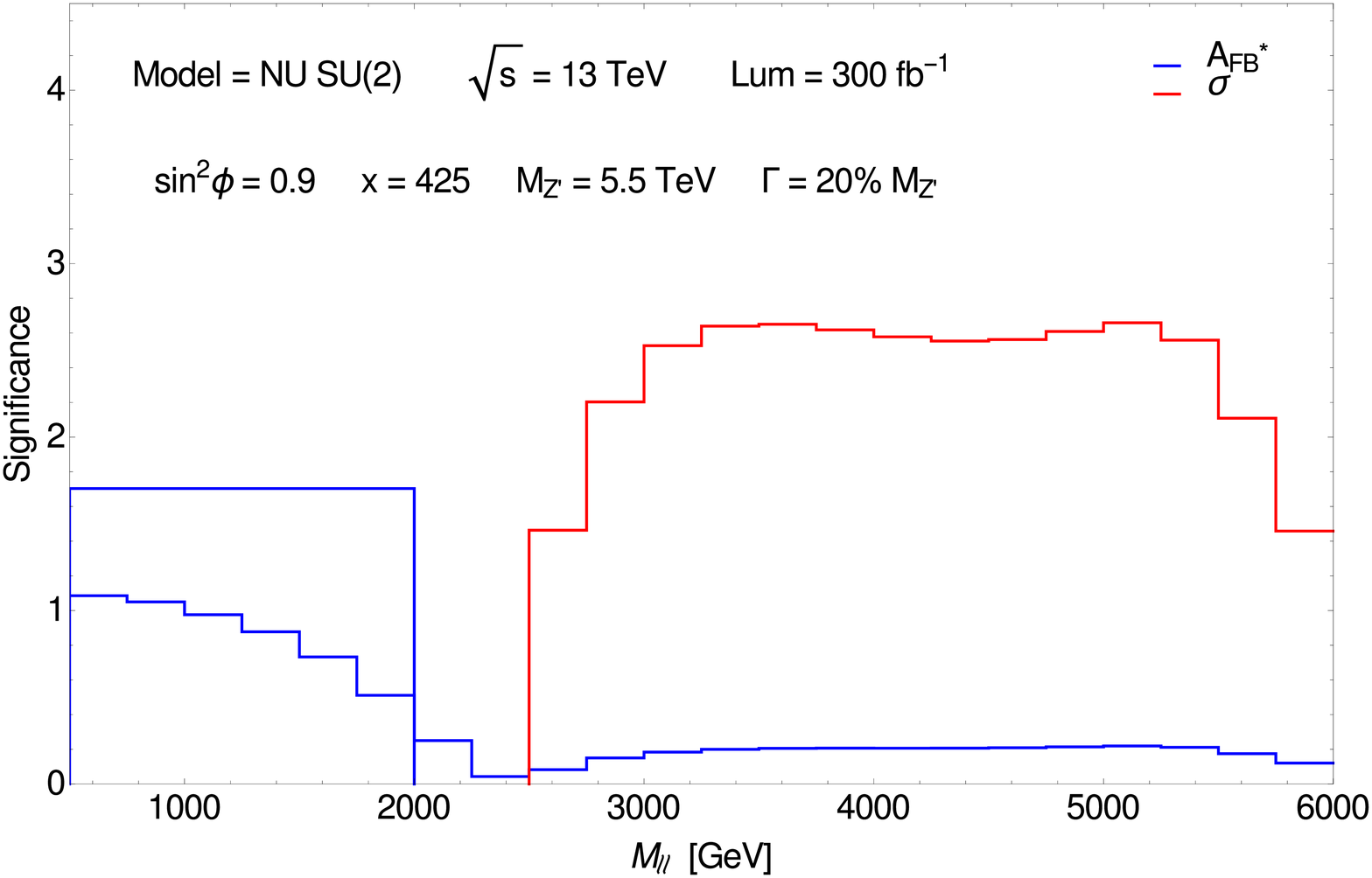}
\caption{Binned significance for the Non-Universal $SU(2)$ model with a $Z^\prime$ with mass $M_ {Z^\prime}$ = 5.5 TeV and $\Gamma_{Z^\prime}/M_{Z^\prime}=20\%$. The results are for the LHC at $\sqrt{s}$ = 13 TeV and $\mathcal{L}=300~ fb^{-1}$, for the two observables: $\sigma$  and $A_{FB}^*$.}
\label{fig:NUSU2_Significance}
\end{figure}

In this case, the invariant mass distribution of the two final state leptons does not show a resonant (or peaking) structure around the physical mass of the $Z^\prime$ standing sharply over a smooth background, but just a broad shoulder spread over the SM background. 
This result is plotted in Fig. \ref{fig:NUSU2}, where we consider a $Z^\prime$ with mass $M_ {Z^\prime}$ = 5.5 TeV and width $\Gamma_{Z^\prime}/M_{Z^\prime}= 20 \%$. 
The line shape of the resonance is not well defined but the shape of the $A_{FB}^*$ distribution could help to interpret a possible excess of events. 
The peak in the AFB distribution is very shifted towards the lower invariant mass region with respect to the $Z^\prime$ pole. In terms of significance of the signal, 
this  translates into an interval with a high AFB significance  for small $\sqrt{\hat{s}}\equiv M_{ll}$  (Fig. \ref{fig:NUSU2_Significance}), that appears much before the cross section excess, although the latter is somewhat stronger.

The experimental method based on the counting experiment assumes that the control region is new physics free. But this might not the case for wide $Z^\prime$s. 
In these scenarios, the interference between the extra $Z^\prime$ and the SM $\gamma , Z$ can be sizable enough to invade the control region. 
If not correctly interpreted, these interference effects could induce an underestimation of the SM background with the consequence of overestimating the extracted mass bounds.  
Having all these uncertainties to deal with, the support of a second observable like AFB is crucial for wide $Z^\prime$ searches.

\section{Conclusions}
In this paper we have considered the scope of using AFB in 
$Z^\prime$ searches at the LHC in the neutral DY channel.
Such a variable has traditionally been used for diagnostic purposes in presence of a potential signal previously established
through a standard resonance search via the cross section. However, based on the observation that it is affected 
by systematics less than cross section, we
have studied the possibility of using AFB as an additional search tool for a variety of $Z^\prime$ models, $E_6$, $GLR$, $GSM$,
embedding either a narrow or wide resonance. The focus was on determining whether such a resonance could be sufficiently wide and/or weakly coupled such that  usual peak searches may not fully identify it and, further, whether the 
AFB could then provide a signal of comparable or higher significance to complement or even
surpass the scope of  traditional analyses. 
We have found promising results. In the case of narrow  $Z^\prime$s, we have proven that 
the significance of AFB based searches can be comparable with the usual bump hunt. 
In the case of wide $Z^\prime$s, the AFB search could have again a comparable sensitivity to the cross section studies 
thanks to a more peculiar line-shape. Furthermore, we have
emphasized the fact that the AFB distribution mapped in di-lepton invariant mass can present features amenable
to experimental investigation not only in the peak region but also significantly away from the latter. 
In essence, here, AFB in specific regions of the invariant mass of the reconstructed $Z^\prime$ could be sensitive to broad resonances much more than the cross section, wherein the  signal seemingly merges with the background.\\

\noindent
{\bf Acknowledgements}~
This work is supported  by the Science and Technology Facilities Council grant  ST/L000296/1.
All authors acknowledge partial financial support through the NExT Institute.

\end{document}